\documentclass[letterpaper,twocolumn,10pt]{article}
\usepackage{usenix}

\usepackage[utf8]{inputenc}

\usepackage{array}
\newcolumntype{?}{!{\vrule width 2pt}}
\usepackage{subcaption}
\usepackage{xcolor}
\usepackage[binary-units]{siunitx}
\usepackage{booktabs}
\usepackage{makecell}
\usepackage{multirow}

\usepackage{authblk}

\usepackage{calc}
\usepackage{xspace}
\usepackage{tikz}
\usetikzlibrary{calc}
\usetikzlibrary{fit}
\usetikzlibrary{positioning}
\usetikzlibrary{shapes.symbols}

\PassOptionsToPackage{
	sortcites,
	maxcitenames=2,
	maxbibnames=8,
}{biblatex}
\usepackage{biblatex}
\bibliography{paper}

\DeclareSIUnit{\nothing}{\relax}
\DeclareSIUnit{\x}{\times}

\newcommand{\Boxer}{{Boxer}\xspace}

\usepackage{authblk}

\begin{document}
\date{}
\title{Short-lived Datacenters}

\author[1]{Michael Wawrzoniak}
\author[1]{Ingo M{\"u}ller}
\author[2]{Rodrigo Bruno}
\author[1]{Ana Klimovic}
\author[1]{Gustavo Alonso}
\affil[1]{Systems Group, Dept. of Computer Science, ETH Z{\"u}rich}
\affil[ ]{\texttt{\normalsize\{michal.wawrzoniak,ingo.mueller,aklimovic,alonso\}@inf.ethz.ch}}
\affil[2]{ INESC-ID / T\'ecnico, ULisboa Lisbon, Portugal}
\affil[ ]{\texttt{\normalsize rodrigo.bruno@tecnico.ulisboa.pt}}

\maketitle

\begin{abstract}
Serverless platforms have attracted attention due to their promise of elasticity, low cost, and fast deployment. Instead of using a fixed virtual machine (VM) infrastructure, which can incur considerable costs to operate and run, serverless platforms support short computations, triggered on demand, with cost proportional to fine-grain function execution time. 
However, serverless platforms offer a restricted execution environment. For example, functions have limited execution times, limited resources, and no support for networking between functions. 
In this paper, we explore what it takes to treat serverless platforms as short-lived, general purpose datacenters which can execute unmodified existing applications. 
As a first step in this quest, we have developed \Boxer, a system providing an execution environment on top of existing functions-as-a-service platforms that allows users to seamlessly migrate conventional VM-based cloud services to serverless platforms.  \Boxer allows generic applications to benefit from the fine-grain elasticity of serverless platforms without having to modify applications to adopt a restrictive event-triggered programming model or orchestrate auxiliary systems for data communication. We implement \Boxer on top of AWS Lambda and extend it to transparently provide standard network interfaces. We describe its implementation and demonstrate how it can be used to run off-the-shelf cloud applications with a degree of fine-grained elasticity not available on traditional VM-based platforms. 
\end{abstract}



\section{Introduction}

Serverless computing platforms such as AWS Lambda~\cite{awslambda}, Google Cloud Functions~\cite{googlefunctions}, and Azure Functions~\cite{azurefunctions} offer highly elastic computing and fine-grained resource billing (as low as milliseconds~\cite{aws-1ms-billing}), while raising the level of abstraction for user interaction with the cloud \cite{Berkeley-CACM}. 
Although serverless platforms cost more per hour than resources on virtual machines \cite{RiseofServerless19}, the fine-grained cost model still makes serverless computing cost-effective for jobs with short execution times and sporadic invocation patterns~\cite{ExCamera,starling,Lambada}. 

Today's serverless offerings are, however, impractical for a wide variety of applications \cite{HellersteinCIDR19,gg,InfiniCache,starling,Lambada}. Current systems limit the execution time of tasks; some restrict the rate at which functions can be created \cite{azurefunctions} or the number of concurrent functions~\cite{Amazon-Lambda-reservation-system}; and none of the platforms support heterogeneous hardware~\cite{aws-FAQs}. 
However, the most critical limitation is the lack of direct communication among functions which has wide-ranging implications for application design and performance. Current attempts at generalizing commercial serverless platforms implement inter-function communication by exchanging data through remote storage systems such as Amazon S3, Anna, and Pocket~\cite{S3, anna, pocket}. Doing so adds considerable overhead and also increases cost~\cite{starling,Lambada}. For instance, parallelism can be used to hide the latency overhead of exchanging data through Amazon S3, with Perron et al. reporting an S3 latency of 14ms to read 256KB of data~\cite{starling}. The consequences are higher costs (each S3 GET and PUT request incurs cost) and a more complex system design which requires parallel reads to a storage system and additional network round trip times. 

In this paper, we explore how serverless computing platforms can be leveraged to run generic applications. Our goal is to provide a general-purpose solution that turns a serverless function into the equivalent of the Linux execution environment that most cloud applications expect. We identify networking as a key challenge, since generic cloud applications rely on direct communication between nodes while serverless platforms do not support it. 
We propose \Boxer, a system that provides a familiar Linux execution environment on top of functions-as-a-service (FaaS) platforms. \Boxer exposes a  Linux socket API for networking inside functions and supports direct, transparent, TCP/IP communication. \Boxer uses a NAT-punching technique to provide bi-directional TCP/IP communication to functions without the external proxies required by existing work ~\cite{ExCamera,gg,InfiniCache} or specializing to a single use case \cite{Boxer-CIDR21}. 

In addition to the design of \Boxer, we make several contributions. First, we demonstrate that \Boxer can run applications that require inter-function communication on AWS Lambda and achieve performance comparable to using similarly sized VMs (\S~\ref{sec:microbenchmarks}). Second, since \Boxer provides the application with a standard network interface, we show in \S~\ref{eval:deathstar} that it supports running unmodified applications, such as the DeathStarBench microservice benchmark suite\cite{DeathStarBench}. Third, we show that \Boxer allows cloud users to leverage the fine-grain elasticity of serverless platforms to quickly absorb load bursts in applications and optimize cost by migrating applications seamlessly between VM and serverless environments depending on the load pattern (\S~\ref{sec:dynamic}). 
Through these capabilities, \Boxer opens up many new opportunities for serverless computing (\S~\ref{discuss:opportunities}). 
By providing conventional networking capabilities using standard interfaces, \Boxer enables a general-purpose, on-demand, short-lived datacenter that can run unmodified, distributed applications and is not limited to trivially parallel jobs. We have implemented and evaluated \Boxer on top of AWS Lambda and we plan on making it available as open source.

\section{Background}
 
Leveraging serverless platforms as short-lived, general purpose datacenters requires bridging the gap between the requirements of distributed applications and the functionality provided by today's serverless platforms. To provide the necessary background, we describe the execution environment of serverless platforms, focusing on AWS Lambda as the most advanced serverless platform in the market. 

\subsection{Serverless Execution Environments}
\textbf{Packaging and invoking functions:} To run a serverless function, users register the function and specify its deployment package, invocation triggers, and memory resource allocation. For the deployment package, AWS Lambda allows users to supply source code or application binaries.
AWS Lambda also supports container images of up to 10 GB in size~\cite{aws-lambda-containers}. In addition to the deployment package, users also configure events to trigger function invocations. Examples of event sources include storage and message queue service notifications, timers, and HTTP requests. Finally, users specify a memory requirement for the function. The cloud provider allocates CPU resources for the function proportionally to its memory allocation. As of October~2021, AWS Lambda supports functions with 128 MB to 10 GB of memory and up to 6 vCPUs (vitual CPUs). Every invocation of a function is executed in its own isolated environment, running a kernel derived from Linux 4.14 and a file system image based on the Amazon Linux~\cite{amazon-linux} distribution. Serverless functions execute as unprivileged userspace processes with restricted capabilities in a secure and isolated runtime environment.

\textbf{Dynamically loading dependencies:} When a function process is invoked and begins executing, a dynamic linker in the AWS Lambda execution environment loads the shared libraries required by the executable. The dynamic linker uses a set of rules to locate the objects that have to be loaded to satisfy the dependency list. It is also possible to direct the dynamic linker to load additional shared libraries before any other shared libraries are loaded. Objects and functions exported by the additional libraries can be then used by the executable. 
If the exported names are the same as the exported names from other libraries loaded later, the executable will use the function or objects from the additional library that is loaded before others. This provides a mechanism to intercept function calls to dynamically linked system libraries such as libc. We use this mechanism to embed \Boxer's interposition library in the system when deploying an application so that network and other operations are selectively routed through \Boxer.  


\subsection{Serverless Networking} 

AWS Lambda functions are assigned private IP addresses 
from private subnets. They are able to send network traffic to external addresses as the traffic from a function to the public internet is routed through Network Address Translation (NAT) gateways. A NAT gateway forwards the traffic to the destination address but changes the source address to a publicly routable address available to the gateway. The assigned mapping between the internal source address and the externally routable source address is stored in the gateway state. As the network traffic arrives at the destination host, its source address is the address assigned by the NAT gateway. When the destination host sends traffic back to the source address it received the traffic from, the traffic arrives back at the gateway having the previously assigned mapping to an internal destination address. Based on the mapping, the gateway modifies the destination address of the traffic and routes it towards the original private source address assigned to the function. For the function, the traffic appears to come directly from the destination host. This results in AWS Lambda functions being able to transparently access external services and internet resources.

Serverless functions cannot accept connections initiated by external sources or by other functions. AWS Lambda restricts the incoming network traffic for functions to traffic that directly corresponds to previously outgoing traffic. Even if an external host learned the private IP address assigned to a function, it could not route traffic to it. Network traffic destined for an internal address must be routed through an external gateway that will then appropriately route it through the private network. If the external gateway is a NAT, then just knowing the route is also not sufficient. A NAT gateway will drop any incoming traffic not matching its previously configured address translation map. Hence, in the case of TCP, a function process can initiate a connection to an external address and then receive traffic on that connection but it cannot accept new incoming TCP connections. 


\section{Related Work}

\subsection{Communication Through Proxies}

Prior work has circumvented the lack of networking by using proxy or coordinator servers to initiate connections and relay messages between serverless functions. Fouladi et al. proposed \textit{mu}, a framework for orchestrating parallel computation and communication across serverless workers, as part of their work on ExCamera~\cite{ExCamera}. The \textit{mu} framework uses a long-lived coordinator and rendezvous server for inter-function communication. Workers running short-lived lambda function invocations establish a connection with the mu coordinator, which can instruct workers to communicate between each other through a rendezvous server. The mu rendezvous server buffers and relays messages from source to destination lambdas. The authors point out that the rendezvous server’s connection to workers can become a bottleneck and recommend implementing direct communication between lambdas via a hole-punching NAT-traversal strategy, but leave this as future work. In a similar approach, Wang et al. propose using serverless functions to build InfiniCache, a distributed in-memory cache~\cite{InfiniCache} that uses a proxy to which functions connect and is used to relay messages to and between serverless functions. A number of other projects, e.g. \cite{gg}, use a similar approach to enable communication. 

Using an fixed infrastructure service that runs parallel to the serverless functions defeats some of purpose of using serverless. Initializing the supporting infrastructure and the proxy/coordinator add overheads, causing starting delays that serverless purportedly removes. Leaving the infrastructure running incurs costs and requires the maintenance, again defeating one of the main goals of serverless.  

We are not the first one to point out the limitations of existing approaches. There was an open-source Serverless Networking SDK that allowed functions to communicate over UDT~\cite{UDT}, a UDP-based protocol (the project is no longer active). It relied on a custom API instead of supporting conventional sockets~\cite{serverlessnet}. Solutions have also been proposed to address the networking overhead of starting thousands of functions at the same time \cite{Particle20} or the overhead of sustaining many RPC calls \cite{Nightcore21}. These solution propose alternative architectures to current commercial deployments and rely on conventional containers or VMs to avoid the limitations of serverless system. In contrast, we are interested in solving these same issues but using the infrastructure provided by cloud vendors. It has been claimed that, in such settings, adding networking makes no sense \cite{Berkeley-CACM} because it would remove the ability of the cloud vendor to optimize the deployment. As \Boxer shows, it is possible to add networking to FaaS without affecting any of its underlying properties or restricting the possibility of optimizing the deployment.  

\subsection{Communication Through Storage}

Applications that require to move larger amounts of data typically resort to an alternative design. Instead of using a coordinator or a proxy, they use ephemeral storage systems, such as Pocket~\cite{pocket}, Anna \cite{Anna18}, and Locus~\cite{Locus}, to improve the performance of inter-function communication through remote storage. In such systems, communication is implemented by writing and reading to a specialized storage layer built for serverless functions. In other cases, rather than using a specially built systems, Amazon's S3 is used directly. For instance, Lambada~\cite{Lambada} and Starling~\cite{starling} are query analytical systems that implement exchange operators as read/write patterns on Amazon S3 as a way to allow functions running parts of a query to exchange data. In both cases, a relatively complex design is needed to deal with the overhead of communication through storage.

While communication through storage solves the problem, it has both a high cost both monetary and in performance. Writing and reading to storage services like S3 is not for free. Implementing communication through storage induces a large amount or read and writes, increasing the cost of using serverless. Performance-wise, using storage to exchange data not only suffers from the higher latency of storage but also adds communication rounds to the exchange. This overhead becomes significantly visible since now AWS Lambda is charged per millisecond and the observed delay is potentially in the order of tens of milliseconds \cite{starling,Boxer-CIDR21}.  

\subsection{General Purpose Serverless}

It has been noted that serverless platforms can be used as a supercomputer on-demand to run highly parallel jobs as it is done in \texttt{gg}, a framework and collection of command line tools to help users run everyday applications -- such as software compilation and unit tests  -- seamlessly on serverless platforms~\cite{gg}. Users express their applications as a composition of lightweight, functional containers using \texttt{gg}'s intermediate representation, while the framework takes care of instantiating containers as serverless functions, loading dependencies, and dealing with function failures and stragglers. This idea of a \textit{supercomputer-by-the-second} also appears in previous wok of the same authors\cite{ExCamera}. Recently, serverless has been presented as the next generation of cloud computing, reinforcing the idea of general purpose use instead of the narrow uses cases considered today \cite{Berkeley-CACM}.  

In this paper we pursue the same notion but focus on generalizing serverless to support unmodified, distributed applications. We aim to turn serverless into a true sub-second datacenter that can be used by off-the-shelf applications rather than just for trivially parallel jobs.

\section{\Boxer}
As a first step towards exploring the notion of the sub-second datacenter, we have developed \Boxer, a system that transparently enables networking between serverless functions. We choose to focus on datacenter applications that use the conventional socket networking API in Linux and the TCP networking protocol, as this represents a large class of datacenter applications. 

\subsection{System Overview}

\Boxer uses two key mechanisms to allow external hosts to initiate connections with serverless functions and functions to communicate with each other transparently. 

First, to allow incoming traffic to be routed to a serverless function, \Boxer uses NAT punching techniques that resemble those proposed in the past for generic NAT services \cite{Eppinger,Ford}. However, our solution is tailored to the AWS Lambda environment and uses functionality specific to AWS Lambda. Through these techniques, \Boxer provides the sending host with the appropriate gateway address and configures the gateway with the appropriate state for the traffic to traverse the NAT and arrive at the private IP address of the destination function. Second, to make the NAT traversal transparent to applications, \Boxer intercepts some dynamically linked libc library function calls by leveraging the dynamic linker that is invoked in the Linux-based AWS Lambda execution environment. For example, when a serverless function calls \texttt{connect}, the \Boxer \texttt{connect} implementation will execute instead of the original libc implementation -- this allows \Boxer to implement NAT traversal without requiring application modifications.

\begin{figure}[t]
  \centering
  \includegraphics[width=.8\linewidth]{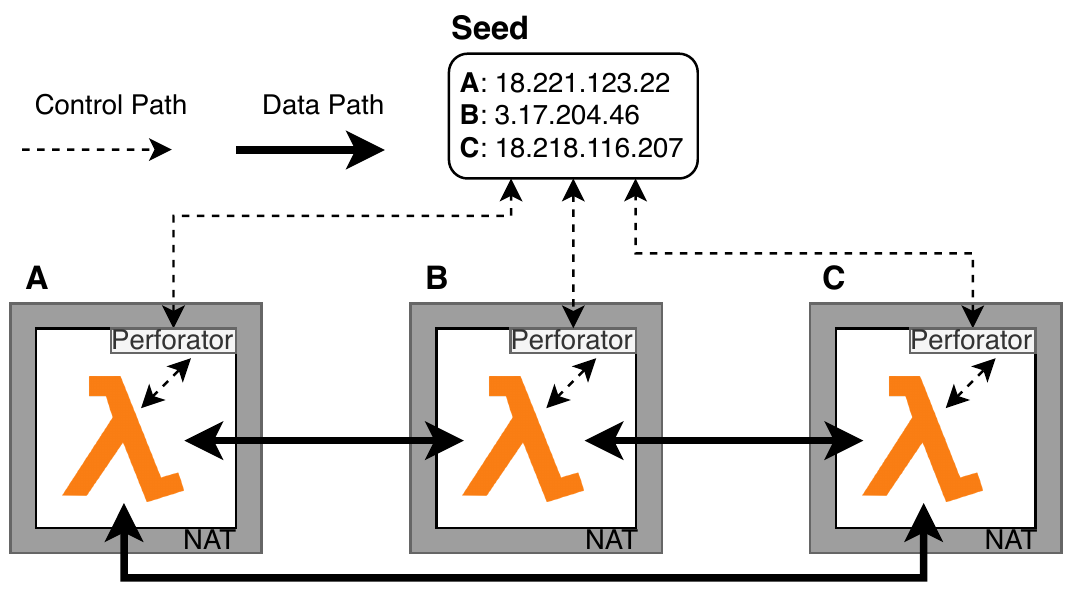}
  \caption{Networked serverless functions use a Seed process to connect functions during the startup. After the startup phase, the seed process is no longer needed.}
  \label{fig:arch}
\end{figure}

\Boxer's networking environment consists of 1) \Boxer processes running inside each serverless function and each node that needs to initiate connections to a function, and 2) a \Boxer `seed' process (Figure \ref{fig:arch}), which runs on an externally routable host and is used for network initialization. Each \Boxer process consists of a \textit{networking service} to establishes TCP connections, a \textit{coordination service} to send and receive network membership updates from other \Boxer processes, and a \textit{transparent execution service} to selectively intercept library calls and run unmodified Linux TCP sockets-based applications. We describe each of these services in more detail below.

\Boxer is packaged as a Linux executable and a shared library object, as we describe in Section~\ref{sec:boxer-packaging}. Users include the \Boxer package in each function deployment package when registering functions on the cloud platform. This allows a \Boxer process to start executing in the serverless function environment when the function is invoked. Users are also responsible for spinning up a \Boxer seed process on an externally routable host, such as an EC2 instance. 




\subsection{Networking Service}
\label{sec:Boxer-net}

The networking service in \Boxer  provides TCP connectivity and manages NAT configuration. We start by describing how nodes join the network and then describe how TCP connections are established.

\begin{figure}[t]
  \centering
  \includegraphics[width=.6\linewidth]{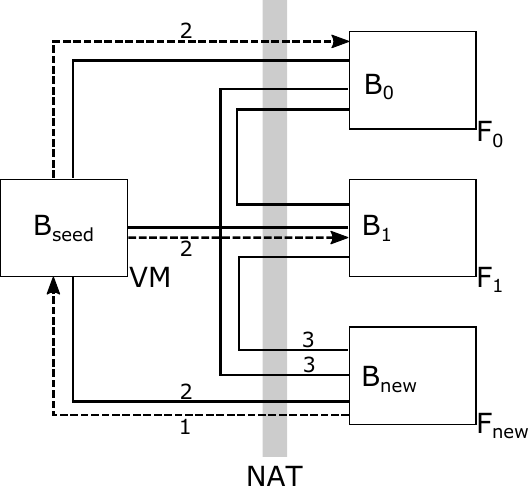}
  \caption{New \Boxer node $B_{new}$ in function $F_{new}$ joins already connected nodes $B_0$, $B_1$ running in functions $F_{0}$,${F_1}$ and seed node $B_{seed}$ running in a virtual machine $VM$. Solid lines represent control connections, dashed lines messages sent, numerical labels the stage of the join protocol when message is sent or connection is established.}
  \label{fig:join}
\end{figure}

\textbf{Joining the network:}
The \Boxer seed process is responsible for network initialization (Figure \ref{fig:join}). The user sets a parameter to specify the address of the seed process. The seed can be any other \Boxer process that is reachable by the process being started. Since functions cannot establish connections to other functions before \Boxer is initialized, the seed process must be running outside of AWS Lambda. We currently run it in an AWS EC2 instance, but it could be run anywhere routable from \Boxer processes that are attempting to join the network. 

The seed process performs three functions. First, it informs the connecting process of its observed external address. If the process is connecting from AWS Lambda, this is the external address assigned by the NAT gateway used. Second, the coordinator service of the seed process informs all other \Boxer processes in the network of a new node, and it supplies the joining process with a list of all other nodes that already joined. Third, the seed process ensures that each external address only exists once. This deals with the fact that two concurrent function invocations may be handled by the same machine behind the same NAT gateway and would thus be indistinguishable. If that happens, all but the first node with a given address are rejected from joining the network, so the corresponding \Boxer process exits immediately and all of the function resources are released.

When a new node joins the network, its coordinator service learns the addresses of all other nodes in the network. Concurrently, coordination services of all other nodes in the network are updated with the address of the newly joined node. The \Boxer networking service of every node listens to membership updates and treats these events as signals that it is time to establish direct control connections with the new node. Control connections are used to exchange commands between \Boxer processes. The membership updates contain the expected external address of the \Boxer network control service. At this point, all nodes have enough information to establish new control connections through the NAT. \Boxer nodes establish the control TCP connections between the new node and the rest of the nodes in the network. This procedure is repeated for every joining node resulting in NxN connectivity through TCP connections, one for each pair of nodes in the network.

\begin{figure}[t]
  \centering
  \includegraphics[width=0.85\linewidth]{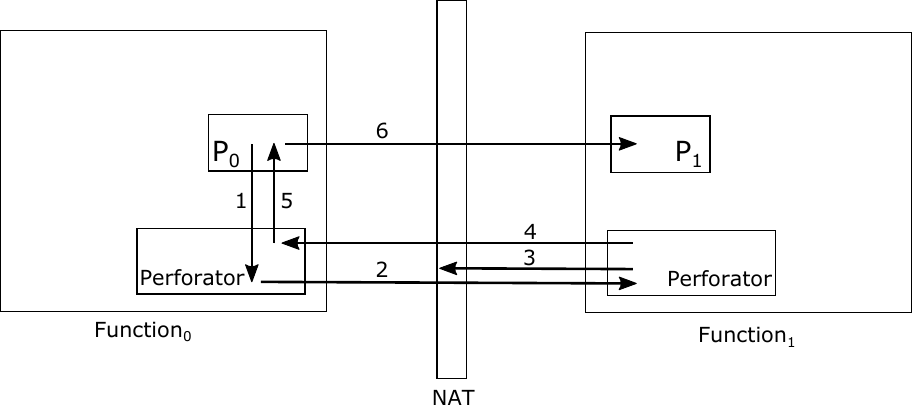}
  \caption{Opening TCP connection by process $P_0$ to process $P_1$ running in a remote function.}
  \label{fig:connect}
\end{figure}

\textbf{Providing TCP connections:}
After the \Boxer networking service is initialized as described above, every node in the network is ready to provide TCP connectivity to its local \Boxer process. The \Boxer networking service provides NAT setup service for its local \Boxer process and to \Boxer processes running on remote nodes. In the current version, the \Boxer networking service does not open TCP connections on behalf of processes; instead, it sets up the appropriate state in the NAT so that when the process attempts to open a TCP connection to another function, the network traffic will traverse through the NAT and the connection can be established.

Figure~\ref{fig:connect} illustrates the TCP connection process. A \Boxer process wishing to establish a TCP connection to a remote \Boxer process cannot immediately open a connection to the remote address because the NAT would block it. Instead, (1) the process sends a request to the local \Boxer networking service to request that the NAT state be configured for the specified connection. If the connection is to an address that is a part of the \Boxer network, then (2) the request is forwarded to the appropriate remote \Boxer network service using the control connections configured during the initialization (described above). The remote \Boxer network service (3) performs the NAT configuration and then (4) sends an acknowledgment back to the requesting \Boxer service that the NAT is ready. (5) The confirmation is forwarded back to the process that made the original request. At this point, the requesting process can (6) attempt opening a TCP connection to the previously specified remote address. Because the remote \Boxer configured the NAT, opening the connection will traverse the NAT and reach the remote process. Assuming the remote process is listening for incoming connections at the specified address, the TCP connection is established.

\subsection{Coordination Service}
Every \Boxer node runs a simple coordination service. The service exposes an interface for local processes to stream membership updates as new nodes join the network. Every membership service instance allows other nodes to become children and propagate the updates. Currently, the seed node is used as the root of the propagation tree. The coordination service is used during the initialization process (described above) by the networking service to establish the control connections. The execution service (described below) uses the coordination service to determine when to proceed with the application execution and to provide the application with the list of peer addresses.

\subsection{Transparent Execution Service}
\label{sec:Boxer-exec}


The \Boxer execution service provides functionality to schedule the execution of applications in a transparent manner. 
To use \Boxer, the programmer specifies a command for \Boxer to run when the \Boxer process starts executing. The command can either be run immediately or after a specified number of members have joined the \Boxer network. For example, \verb|perforator -s SEED_ADDR| \verb|-n 5 zk-start.sh| will run the \verb|zk-start.sh| command once the network has five members.  This mode of execution is particularly useful to start datacenter applications that require a static network configuration.
If a barrier for a target number of workers is specified, \Boxer runs the specified command and sets its environment variables to indicate the local node address, a unique node ID assigned to the current node, and a file name of a file containing the list of addresses for all other nodes. If no barrier is specified and the command runs immediately, \Boxer will still learn of network membership changes during execution via the coordination service. 
To run datacenter applications, users may write small shell scripts that generate the necessary configuration files based on the environment variables and then start the application. 

\begin{figure}[t]
  \centering
  \includegraphics[width=.65\linewidth]{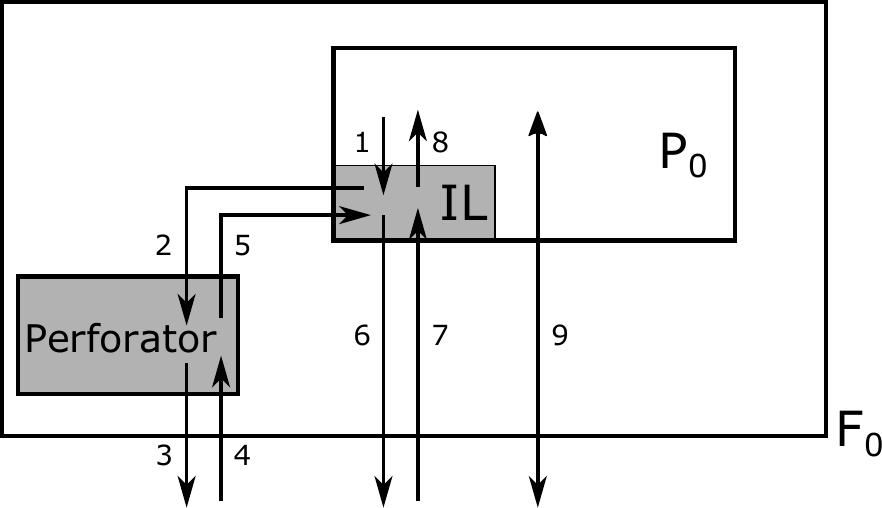}
  \caption{Transparent application execution via the interposition library(IL). Application process $P_0$ uses socket \texttt{connect} function to open a TCP connection to a remote function.}
  \label{fig:interpose}
\end{figure}

\subsection{Interposition Library}

To enable applications to establish connections using \Boxer without requiring changes to application code, when \Boxer runs commands, it instructs the dynamic linker to load the \Boxer interposition library before all other shared libraries are loaded and the application execution begins. The \Boxer interposition library is designed selectively to intercept functions provided by other shared libraries that are loaded by the dynamic linker. 

We design \Boxer to provide transparent networking support for applications that use the socket interface, which is generally provided by a version of the libc system library on Linux systems. Applications use the \verb|connect| function on stream sockets to open TCP connections. When the interposition library is loaded, applications use the \verb|connect| function exported by the interposition library, instead of the default version provided by libc. 
This provides the primary mechanism for intercepting the necessary function calls to provide transparent sockets semantics using \Boxer networking to establish TCP connections.

Figure ~\ref{fig:interpose} illustrates the process of transparently opening remote TCP connections by an application running in a function with \Boxer. Every time an application process calls the \verb|connect| function, it is intercepted by the interposition library (1) If the call is not associated with a TCP socket, the interposition library immediately forwards the call to the default implementation of \verb|connect|. Otherwise, if the socket is not already bound to a local address, it is bound to a local address assigned by the kernel, and then the local and remote addresses associated with the socket are used to construct the appropriate connection request command. The command is then sent to the local \Boxer networking service. (2) The networking service forwards the request to the appropriate remote node (3) and, when the result is received, (4) it reports it back to the library. (5) If a successful acknowledgment is returned from \Boxer networking, that indicates that the NAT is configured, so the interposition library (6) establishes a TCP connection using the default version of the \verb|connect| function. When the system \verb|connect| function completes, (7) the result of the final call is returned to the calling application. (8) The application does not observe any functional difference from the default \verb|connect| call. If the connection was successful, the application continues to use the TCP connection in the same way as if \Boxer was not involved in the process. No function calls for sending and receiving data over the socket are intercepted; instead, (9) default library functions are used directly.

In addition to the \verb|connect| function, to provide transparency to the applications accepting TCP connections, the \verb|bind| socket function is also intercepted by the interposition library. The NAT configuration process performed by the networking service requires binding a temporary socket to the same local address as the socket that an application accepts connections on. To make this possible, the application's listening socket must have the \verb|SO_REUSEPORT| socket option set. To support unmodified applications, \Boxer modifies this socket option for the application by intercepting \verb|bind| calls (the same effect could be achieved by intercepting other functions as well). It should be noted that this is an example where perfect transparency is not preserved. The application can detect that the option was changed on the socket, and if the application relies on this option not being set, then it could have an impact on the application. However, we do not aim to cover unusual corner cases, and typical datacenter applications are functionally not affected.

\begin{table*}[h!]
\centering
\small
\resizebox{\textwidth}{!}{%
\begin{tabular}{|c|c|c|c|c|c|c|c|c|c|c|}
\hline
\multirow{2}{*}{TCP connection type} & \multicolumn{2}{c|}{Mean} & \multicolumn{2}{c|}{Median} & \multicolumn{2}{c|}{Std.} & \multicolumn{2}{c|}{Min} & \multicolumn{2}{c|}{Max}
\\
\cline{2-11}
& Forward & Reverse & Forward & Reverse & Forward & Reverse & Forward & Reverse & Forward & Reverse
\\
\hline
Function-to-Function      & 622.57 & 622.63      & 626.88 & 628.59      & 24.32 & 25.81        & 564.25 & 561.16      & 680.63 & 678.36      \\ 
VM-to-Function            & 428.28 & 426.03      & 429.03 & 428.09      & 1.69 & 4.01          & 420.31 & 415.62      & 429.16 & 429.18      \\ 
Function-to-VM            & 410.31 & 427.05      & 422.64 & 428.74      & 26.58 & 3.61         & 335.70 & 414.39      & 429.06 & 430.33      \\ 
Function-to-VM-native   & 427.77 & 426.67      & 429.04 & 428.62      & 3.16 & 4.93          & 412.74 & 399.35      & 429.07 & 429.07      \\ 
VM-to-VM                  & 428.89 & 428.96      & 428.96 & 428.95      & 0.55 & 0.29          & 424.78 & 427.50      & 429.01 & 430.39      \\ 
VM-to-VM-native         & 429.02 & 429.06      & 429.03 & 429.07      & 0.15 & 0.02          & 428.43 & 429.02      & 429.65 & 429.09      \\ 

\hline
\end{tabular}}
\caption{TCP throughput in Mbit/s for different connection types measured using iperf3\cite{iperf3} 
Forward and Reverse modes refer to throughput achieved by the client side generating traffic or server side respectively.VMs are EC2 m4.large instances.}
\label{fig:throughput}
\end{table*}

\begin{table*}[h!]
\centering
\small
\resizebox{\textwidth}{!}{%
\begin{tabular}{|c|c|c|c|c|c|c|c|c|c|c|}
\hline
\multirow{2}{*}{TCP connection type} & \multicolumn{2}{c|}{Mean} & \multicolumn{2}{c|}{Median} & \multicolumn{2}{c|}{Std.} & \multicolumn{2}{c|}{Min} & \multicolumn{2}{c|}{Max}
\\
\cline{2-11}
& Forward & Reverse & Forward & Reverse & Forward & Reverse & Forward & Reverse & Forward & Reverse
\\
\hline
Function-to-Function      & 621.48 & 621.44      & 628.42 & 627.95      & 23.87 & 24.28        & 560.61 & 560.99      & 674.14 & 677.28      \\ 
VM-to-Function            & 622.98 & 624.10      & 610.70 & 623.97      & 26.59 & 32.31        & 595.92 & 566.07      & 680.64 & 681.22      \\ 
Function-to-VM            & 623.18 & 621.78      & 637.01 & 640.82      & 26.99 & 32.22        & 564.00 & 570.34      & 658.91 & 664.03      \\ 
Function-to-VM-native   & 624.13 & 622.39      & 639.62 & 622.06      & 27.75 & 31.24        & 566.61 & 571.08      & 657.34 & 668.00      \\ 
VM to VM                  & 4684.53 & 4706.22    & 4744.35 & 4735.99    & 144.29 & 103.36      & 4034.14 & 4263.13    & 4789.13 & 4787.11    \\ 
VM-to-VM-native         & 4770.42 & 4695.97    & 4786.23 & 4740.02    & 39.46 & 159.41       & 4571.73 & 3903.98    & 4787.20 & 4789.87    \\ 

\hline
\end{tabular}}
\caption{TCP throughput in Mbit/s for different connection types measured using iperf3\cite{iperf3} between 32 pairs of hosts for each type. Connection types are the same as in Table~\ref{fig:throughput}. 
The measurements were conducted in AWS eu-west-3 region, VMs are EC2 m5.large instances.
}
\label{fig:throughput_big}
\end{table*}

\subsection{Packaging Applications with \Boxer}
\label{sec:boxer-packaging}

In order to run existing applications in serverless functions,
they have to be packaged together with \Boxer
using the deployment process of the cloud provider.
A typical application package consists of four components:
(1) the event handler the function service calls when the function is invoked,
(2) the \Boxer executable and interception library,
(3) the application itself including templates of its configuration files,
and (4) a dynamic configuration script.
In AWS, we have used dependency layers for the \Boxer and application code
and deployed the handler and configuration scripts as the function code. \Boxer introduces a small function deployment package size overhead resulting from the \Boxer executable and interception library. These components account for 3.9 MB and 2.8 MB before compression, and 873 KB and 545 KB after compression (\Boxer and interception library, respectively). In total, the package size increases by 1.4 MB, a reasonable overhead considering the current deployment package size limit of 50 MB.

The event handler typically starts the networking service
such that the newly invoked function joins the network.
The address of the seed process, which is required for that process,
can be passed as a function parameter in the invocation
or read from some pre-defined place in the cloud.
The execution service then starts the dynamic configuration scripts;
depending on the selected start-up mode, this happens
either immediately or after enough nodes have reached the barrier.
The configuration script consults the coordination service
for the current list of nodes in the network and their addresses
and generates the runtime configuration of the application
using the configuration template based on that list.
When everything is set up, the script runs the application itself.

\section{Benchmarking \Boxer}\label{sec:microbenchmarks}
To evaluate \Boxer in practice, we start by characterizing its throughput and latency. 

\subsection{Throughput analysis}\label{eval:microbenchmarks}
We summarize the networking characteristics we observe in AWS Lambda using \Boxer. We show what a typical application running in AWS Lambda with \Boxer can achieve in terms of TCP throughput and latency. Unless noted otherwise, all measurements were performed in AWS us-west-2 region with Lambda functions with 3008MB of memory and m4.large EC2 VM instances.

\Boxer enables TCP connectivity between AWS Lambda serverless functions and allows outside hosts, running outside of AWS Lambda to initiate TCP connections to serverless functions. As described in Section~\ref{sec:Boxer-net} this is achieved by traversing the NAT gateways between hosts. 
During development we have seen unfavorable network conditions when \Boxer used different methods to establish connectivity. This is because the cloud applications we want to enable usually expect symmetric network properties between end-points. However, given that \Boxer traverses an unknown network of middle-boxes that may impose arbitrary network filtering or throttling rules, we must verify the properties of the various scenarios the middle-boxes could differentiate, based on ordering of packets, timing or types of end-hosts. 
Thus, in the evaluation we distinguish six connection types; (1) Function-to-Function connections are established by \Boxer between a pair of AWS Lambdas, (2) VM-to-Function are connections initiated by \Boxer running in an EC2 VM to an AWS Lambda also running \Boxer, (3) Function-to-VM are initiated by \Boxer running in AWS Lambda to a EC2 VM also running \Boxer, (4) Function-to-VM-native are connections initiated from AWS Lambda to EC2 VM without the use of \Boxer (this scenario is allowed by default on AWS Lambda) as it is used a baseline, (5) VM-to-VM are connections established using \Boxer between a pair of EC2 VMs and (6.) And as a baseline, VM-to-VM-native are vanilla connections established between a pair of EC2 VMs without \Boxer. The evaluation also proves the versatility of \Boxer and the many configurations in which it can be deployed.


\begin{table*}[h!]
\centering
\small
\begin{tabular}{|c|c|c|c|c|c|c|c|c|c|c|}

\hline
\multirow{2}{*}{TCP connection type} & \multicolumn{5}{c|}{Round-trip latency of 1k byte message ($\mu s$)} & \multicolumn{5}{c|}{Connection establishment time for time-to-first byte ($\mu s$)}
\\
\cline{2-11}
& Mean & Median & Std. & Min & Max & Mean & Median & Std. & Min & Max
\\
\hline

Function-to-Function      & 694.23               & 758.00               & 289.52               & 202.00               & 2769.00 
                          & 2735.21              & 2625.00              & 10001.00             & 890.00               & 1033112.00\\ 
VM-to-Function            & 547.84               & 457.00               & 194.29               & 244.00               & 2471.00
                          & 1981.38              & 2153.00              & 7909.74              & 821.00               & 1011171.00\\ 
Function-to-VM            & 520.10               & 436.00               & 189.53               & 244.00               & 2372.00
                          & 2086.03              & 2239.00              & 6124.57              & 882.00               & 1015205.00\\ 
Function-to-VM-native     & 622.53               & 686.00               & 175.57               & 241.00               & 2434.00
                          & 1378.56              & 1244.00              & 8027.04              & 382.00               & 1012935.00\\ 
VM-to-VM                  & 193.69               & 188.00               & 43.34                & 150.00               & 2165.00
                          & 1067.24              & 1034.00              & 166.09               & 894.00               & 7402.00\\ 
VM-to-VM-native           & 197.62               & 194.00               & 28.18                & 153.00               & 1862.00
                          & 407.81               & 345.00               & 284.37               & 258.00               & 6416.00\\ 

\hline
\end{tabular}
\caption{Round-trip latency and connection establishment times for different TCP connection types.}
\label{fig:latency}
\label{fig:ttfb}
\end{table*}

We benchmark TCP throughput of different connection types between pairs of hosts (VMs or functions) by running unmodified iperf3\cite{iperf3} tool as a \Boxer application (or natively for the native connection types). We instantiate 32 non-overlapping pairs of functions for a 60 seconds period for each scenario. In each pair, one function runs iperf3 in server mode, and one in client mode. The iperf3 client function connects to the listening server to begin the configured benchmark. When configured in the forward mode, the client side generates TCP traffic, when configured in the reverse mode, the server side generated the TCP traffic. We measure both to verify that the underlying network does not apply different network policies in different directions (we have seen this during development.) The achieved throughput reported on the receiving side at 1 second interval. 

Table~\ref{fig:throughput} presents throughput statistics for different connection scenarios. The sustained average throughput is 622Mbit/s in forward and reverse direction between a pair of AWS Lambda functions running \Boxer (Function-to-Function). The variance level is low; the throughput is steady and sustained throughout the connection. In an additional experiment, we verified that the throughput can be sustained throughout the maximum lifetime of an AWS Lambda function (currently 15 minutes).
The observed throughput between a pair of VMs is 429Mbit/s if \Boxer is used or not, demonstrating that \Boxer adds no data-plane overhead (after a connection is established). Throughput between functions and VMs is similar (410-428Mbit/s) and symmetric in all connection scenarios. In this case it is limited by the throughput of the VMs (m4.large) network. Table~\ref{fig:throughput_big} shows the same benchmarks but using a higher-bandwidth VM network (m5.large instances): the upper-bound on the throughput between AWS Lambda and VMs is the same as the throughput of the AWS Lambda internal network of 621Mbit/s (the experiment is performed in a different AWS region, and AWS Lambda function-to-function throughput is the same).
The achieved TCP throughput between VMs and functions matches that observed by others ~\cite{Lambada} between functions and AWS services such as S3.

To gain further insight about the bandwidth limits enforced, we conduct a load testing experiment by concurrently sending data from multiple functions to one. The server executing in one function listens for connections from clients executing in $N$ other functions. Each client establishes one TCP connection to the server and attempts to saturate the connection by sending data in a tight loop. At 1 second interval, the server records aggregate bytes received from all of the clients. We vary the number of clients from 1 to 256 functions and run each configuration over a 5 minute interval. Figure~\ref{fig:recvs_n} presents the averages of the aggregated received throughput at the server after removing the initial and final 30 seconds of the experiment measurements. The maximum observed throughput is 621.69Mbits/s with 1 sending function and minimum of 607.04Mbit/s at 128 sending functions. We attribute the degradation of less than 3\% to the overhead associated with handling multiple connections. This leads us to the conclusion that the ingress TCP bandwidth limits imposed on AWS Lambda functions do not depend on the number of sending function and that the available bandwidth is comparable to that of regular instances.
 
The only resource parameter that can be adjusted for AWS Lambdas is the amount of memory, which then proportionally determines the vcpu share allocated to the function. To determine if the memory setting also influences network properties we varied memory allocated to functions and measured the achievable throughput. The throughput did not vary with memory settings of 512MB, 3008MB, and 10240MB.


\subsection{Latency analysis}

\begin{figure*}[t]
  \centering
  \begin{minipage}[t]{.3\textwidth}
    \centering
    \includegraphics[width=1.1\linewidth]{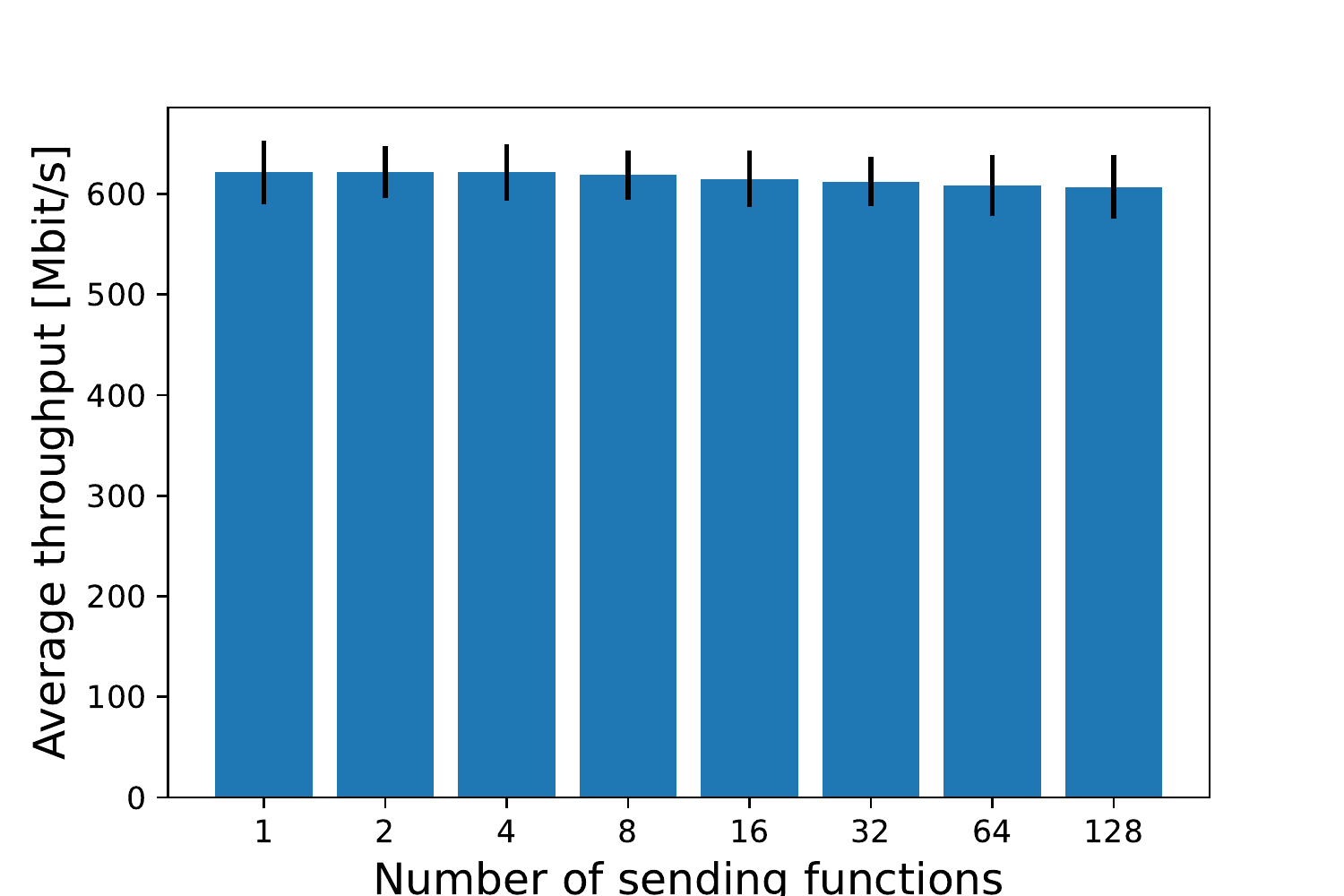}
    \caption{Comparison of aggregate receive throughput as number of sending functions varies. Maximum is 621.69Mbits/s at 1 sending function and minimum at 607.04Mbit/s at 128 sending functions, black lines represent standard deviation.}
    \label{fig:recvs_n}
  \end{minipage}
  \hspace{.5cm}
  \begin{minipage}[t]{.3\textwidth}
    \centering
    \includegraphics[width=1.1\linewidth]{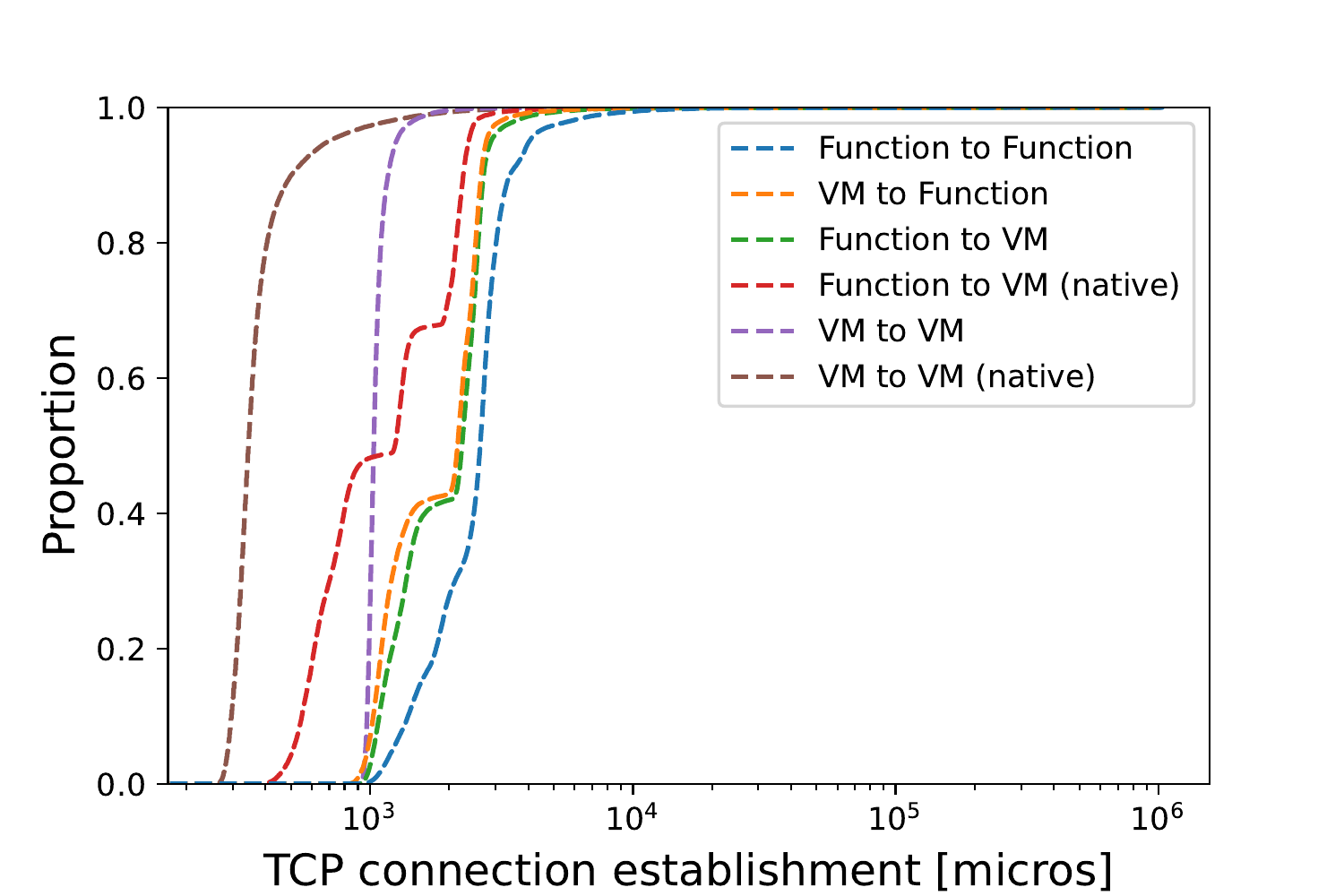}
    \caption{Empirical CDF of TCP connection establishment times. Time-to-first-byte (TTFB) for different connection types measured in microseconds between 32 distinct pairs of hosts establishing 1024 TCP connections each.}
    \label{fig:connect_ecdf}
  \end{minipage}
  \hspace{.5cm}
  \begin{minipage}[t]{.3\textwidth}
    \centering
    \includegraphics[width=1.1\linewidth]{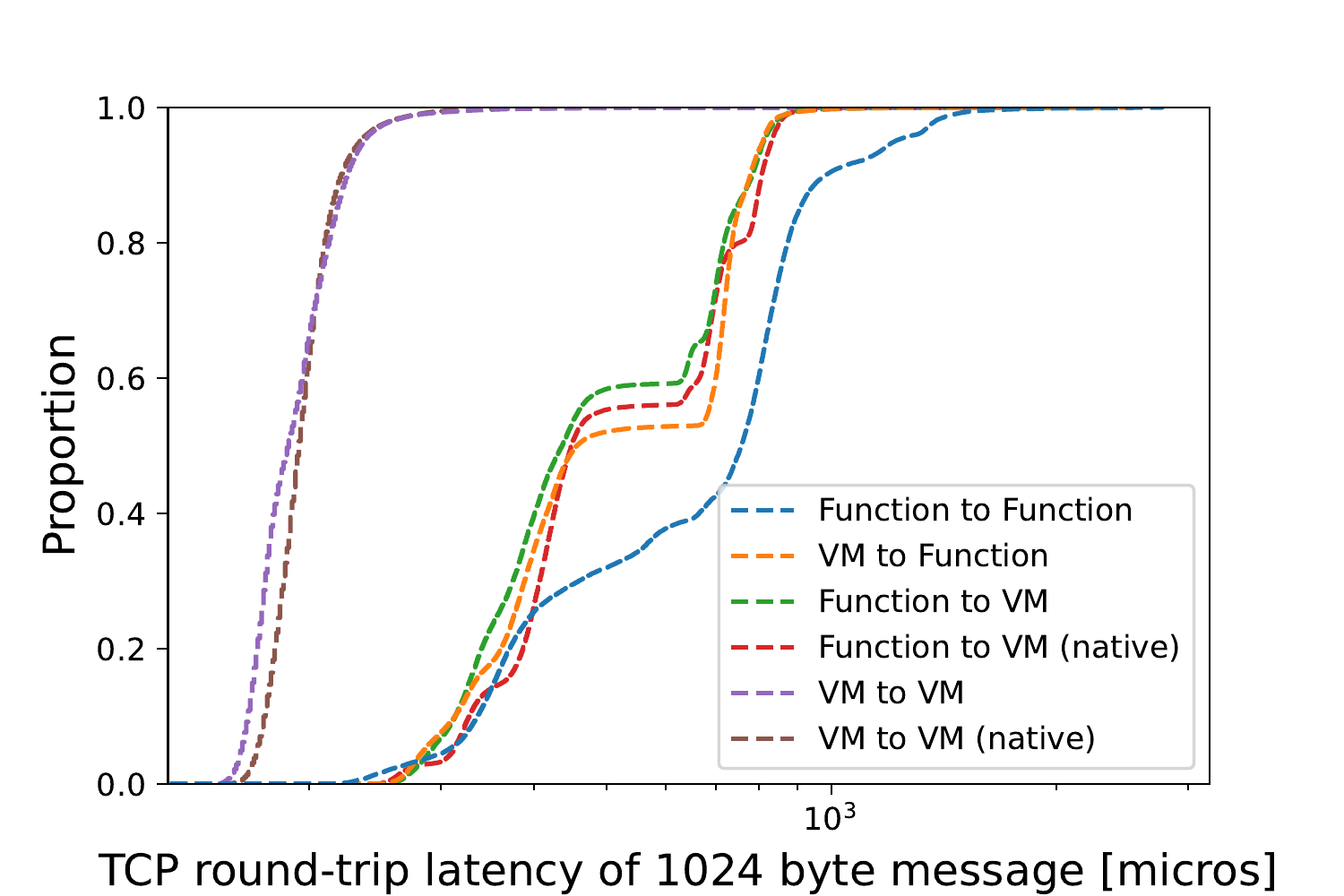}
    \caption{Empirical CDF of TCP round-trip latencies between 32 distinct nodes pairs (of VMs and AWS Lambda functions) echoing 1024 byte message, repeated for 6 connection scenarios.}  
    \label{fig:latency_ecdf}
  \end{minipage}
\end{figure*}



We measure TCP latency of the six connection types described above. For each connection type, we instantiate 32 non-overlapping pairs of hosts (AWS Lambda functions or VMs). Each host executes a benchmarking program, and every pair opens a single TCP connection with Nagle's algorithm disabled by both hosts. The host assigned the client role implements a TCP echo client that initiates a connection to the assigned server host. After accepting the connection from the client, the server function initiates 128 rounds of ping-pong exchanges of a 1024-byte message and measures the total time, this measurement is repeated 1024 times. 
Table~\ref{fig:latency} list the summary statistics of observed round-trip latencies for different connection types and Figure~\ref{fig:latency_ecdf} shows the eCDF of the latencies measured.

\Boxer's TCP connections between pairs of functions have a mean round-trip latency of $694\mu s$. These connections also show significant variability in latencies, with a range between $202\mu s$ and $2769\mu s$, as can be seen in the eCDF plot.
The main source of the observed variance is not due to the variance within each TCP connection but due to the variance between different TCP connections. This suggests that the network between different Lambda functions, from the latency perspective, is not uniform, and different instantiation patterns result in different latencies. This makes sense as the distance between the machines where functions are deployed plays a role in determining the latency.

The latency measurements of the VM-to-VM-native connections, without the use of \Boxer, and VM-to-VM connections that are established by \Boxer show very close round-trip latencies of $198\mu s s$ and $194\mu s$, and follow a very similar distribution which can be seen in the eCDF plots of the two connection types. This further shows that there is no data plane overhead of \Boxer provided connections once they are established.
The observed latency of connections between Lambda functions and VMs have similar mean round-trip latencies ranging from $520\mu s$ to $622\mu s$, and all follow a similar distribution as can be seen in the eCDF plot. Perhaps surprising, the connections provided by \Boxer have a slightly lower round-trip latency, which could be attributed to a different processing of the traffic by the network due to its different initial packet signature (we may investigate this further in the future.)

We observe that the mean round-trip latency of \Boxer's TCP connections between pairs of functions is 3.51$\times$ greater than latencies observed on AWS (m4.large) VM-to-VM native TCP connections. We consider this acceptable given the added functionality of \Boxer. 

\subsection{Connection establishment}
We measured the time required for application to have an established TCP connection for the six connection types described above. For each connection type we instantiate 32 host pairs (AWS VMs or Lambda functions depending on the scenario) and assign one host in each pair to be the client and the other to be the server. The time-to-first-byte(TTFB) measurement is recorded by the client. The client starts a timer, attempts to connect to its server and waits to read 1 byte of data from the server. Once it receives the 1 byte it stops the timer, records the duration and closes the connection. The server accepts connections and replies with 1 byte right after it accepts a new connection. This is repeated 1024 times by each pair, the measurements are in Table~\ref{fig:ttfb}.

Unlike TCP throughput and latency, \Boxer does have overhead compared to native TCP connection times. Mean time to establish a function to function connection is $2735\mu s$. The overhead can be seen in comparison of establishment time for VM-to-VM-native connections of $408\mu s$ vs VM-to-VM connections that are established by \Boxer of $1067\mu s$. This is due to additional round trip necessary to contact the destination \Boxer to request that the connection setup and wait for acknowledgement before proceeding locally, or in case of an error response to forward to the error to the application (for example in case when there is nothing listening at the destination address.) This additional signaling adds to the connection setup times, and can also be observed in the connection times between functions and VMs. However, given that there are no alternative native connections for function-to-function and VM-to-function connection types, we consider the latency acceptable.

The eCDF of the TTFB times for the different scenarios in Figure~\ref{fig:connect_ecdf} shows that the comparable scenarios with and without \Boxer connection setup follow similar distribution, but are shifted by additional delay of the round trip times. Function-to-function connections, and connection types that cross the AWS Lambda and EC2 networks have maximum TTFB times observed to be over 1 second, including a connection scenario that does not involve \Boxer to establish TCP connections (Function-to-VM-native.) This suggests that there may be packet loss or network congestion in the network and it is not a direct consequence of the procedure \Boxer uses to establish the connections. As it can be seed from the eCDF such extreme connection times are rare, in the case of Function-to-Function connection type, 99.9 quantile for TTFB is $20654\mu s$. Compared to communicating through storage, such latencies are acceptable, especially considering that in all of the above experiments performed, all connections were successfully established.









\section{Short-lived data center}
\label{eval:deathstar}
To demonstrate the ability of \Boxer to provide a short-lived datacenter, we run an unmodified, complex distributed application: the DeathStarBench benchmark \cite{DeathStarBench}. DeathStarBench is a suite of cloud microservice benchmarks deployed using container networks which mimic how real applications are deployed in production. We show that \Boxer allows us to transparently deploy stateless microservices using AWS Lambda instead of AWS EC2 (VMs) or Fargarte (Containers). 

\begin{figure*}[t]
  \centering
  \begin{minipage}[t]{.33\textwidth}
    \includegraphics[keepaspectratio,width=\textwidth]{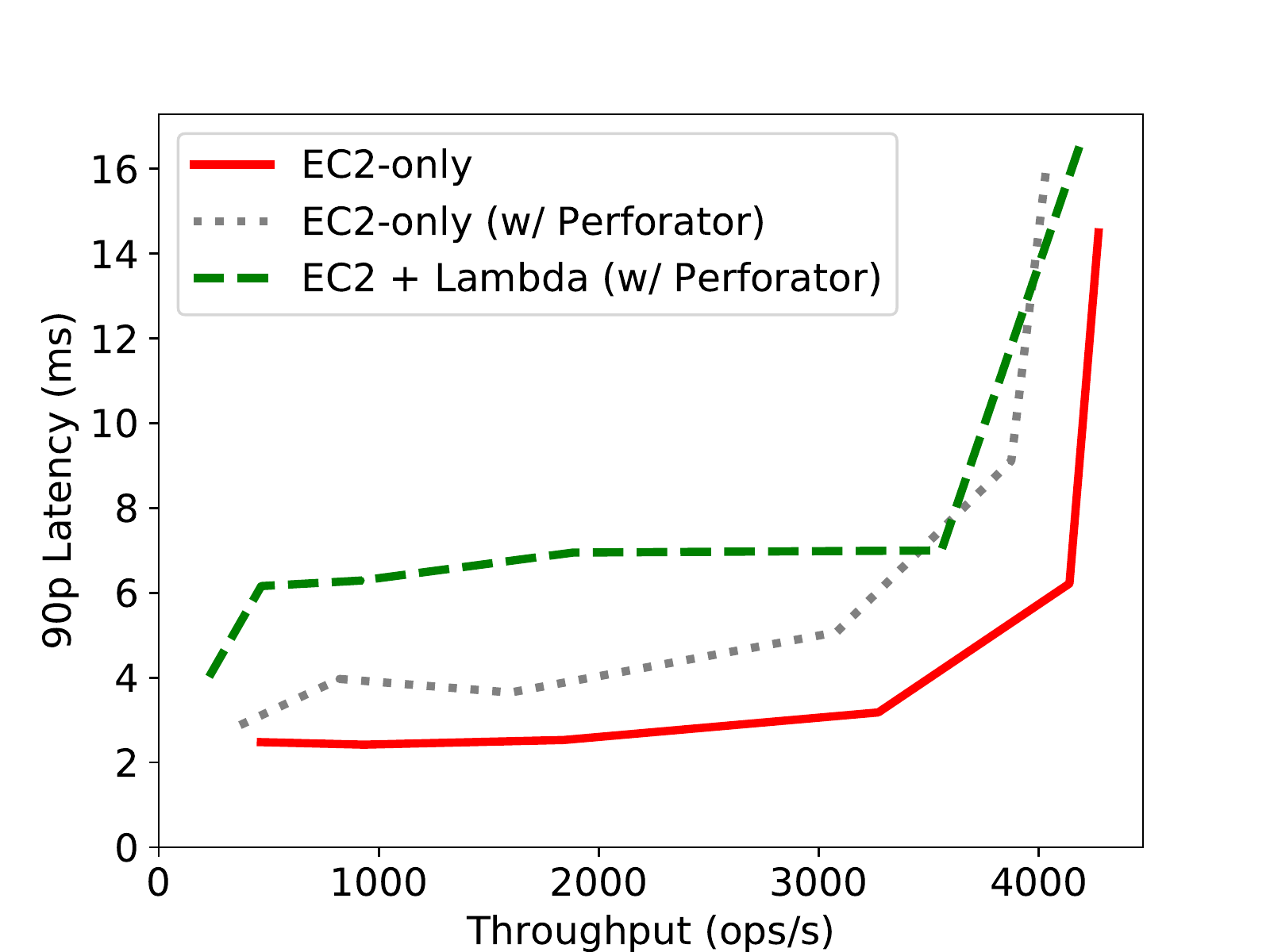}
    \centering
    a) Read Workload
  \end{minipage}                     
  \begin{minipage}[t]{.33\textwidth}
    \includegraphics[keepaspectratio,width=\textwidth]{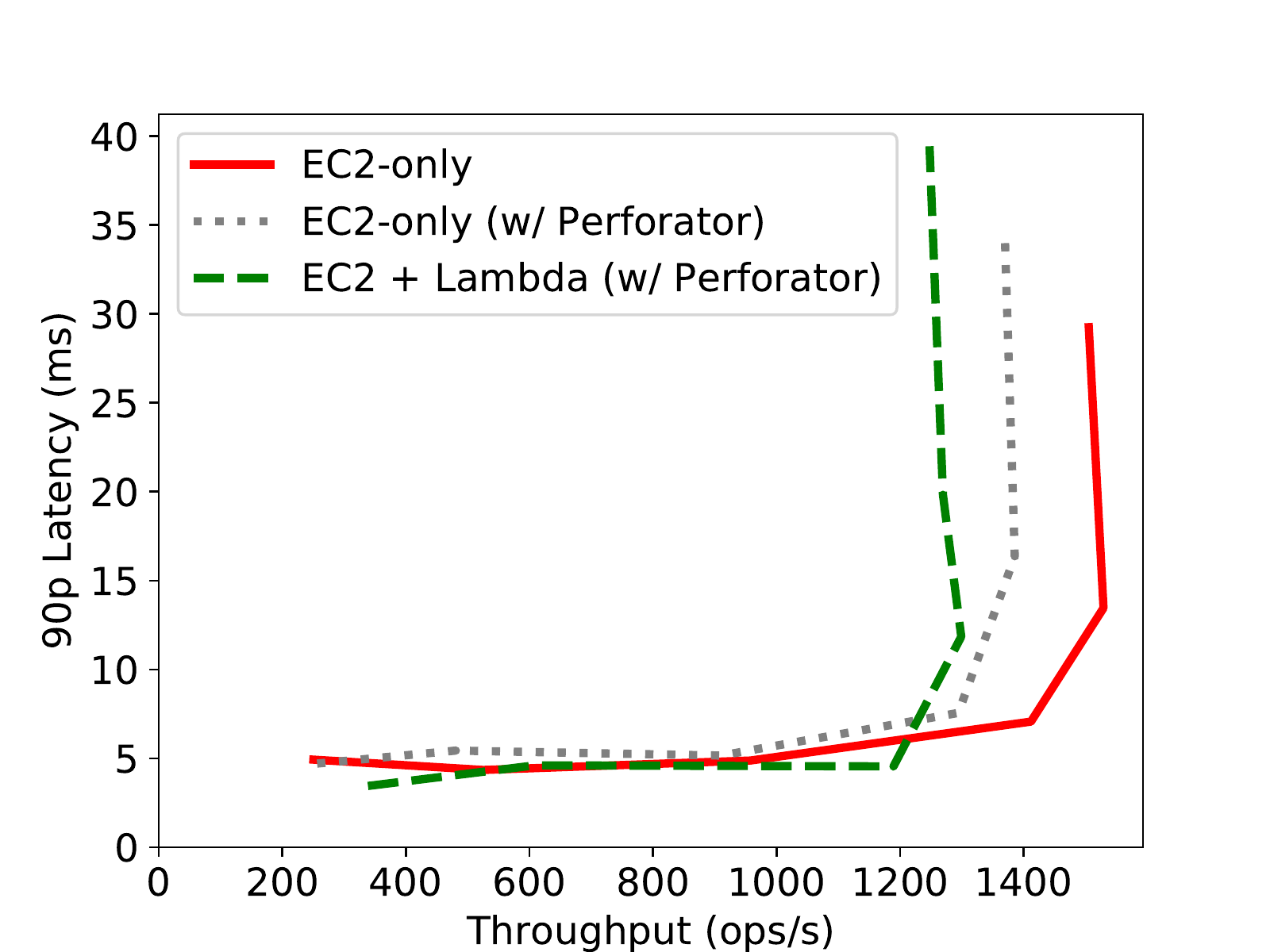}
    \centering
    b) Write Workload
  \end{minipage}
  \begin{minipage}[t]{.33\textwidth}
    \includegraphics[keepaspectratio,width=\linewidth]{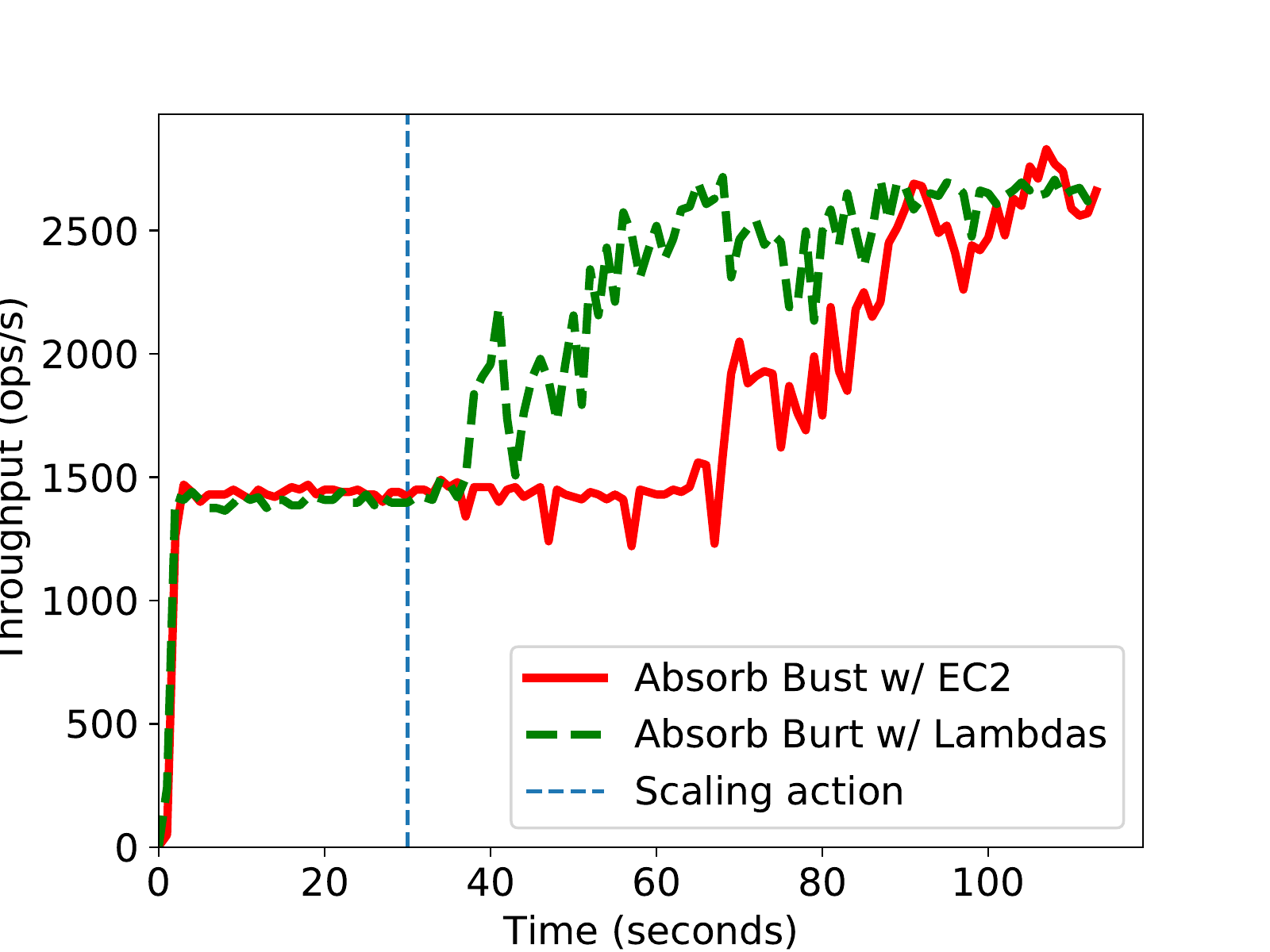}
    \centering
    c) Burstable Write Workload
  \end{minipage}
  \caption{DeathStarBench experimental results for read and write workloads in a static deployment (left and center); burstable write workload in dynamic deployment (right).}
  \label{fig:eval_deathstar}
\end{figure*}

\subsection{Using \Boxer with the DeathStarBench}
We focus on DeathStarBench's \textit{socialNetwork} application, which offers a social network service to users and is organized using three microservice layers: i) front-end layer (implemented using an NGINX webserver); ii) logic layer (implemented using stateless Thrift services that communicate through RPCs); iii) caching and storage layer (implemented with MongoDB and Memcached instances). In  \textit{socialNetwork}, user requests are received by the front-end layer (NGINX web server) and then routed to one of the services in the logic layer. Depending on the user request, the logic layer may perform one or multiple requests to the caching and storage layers. Since the logic layer is stateless (i.e., it contains no internal persistent state), it can be deployed on AWS Lambda. However, since functions cannot receive connections from external components, (e.g., the front-end NGINX laye in this case), \Boxer is required to establish such connections.

We did not have to make any modifications to the application code to deploy DeathStarBench on AWS Lambda with \Boxer. The benchmark was only modified to i) use names instead of fixed local IPs (for example, replace \texttt{127.0.0.1} by \texttt{nginx-thrift}), and ii) wrap the invocation of all components of the front-end and logic layers with the \Boxer binary (for example, replace \texttt{SocialGraphService} with \texttt{\Boxer~-s <seed>} \texttt{-n <number of components>} \texttt{SocialGraphService}). Wrapping the invocation of the front-end and logic layer components with \Boxer ensures that the creation of new connections goes through the \Boxer network, allowing not only the front-end layer to establish connections to services running in the logic layer (VM to Lambda), but also components in the logic layer to establish connections between them (Lambda to Lambda).

\subsection{Methodology}
To evaluate the performance impact of moving the logic layer to Lambda using \Boxer, we use three types of deployments. First, an \textit{EC2-only} deployment in which all components are deployed as VMs in EC2 is used as our baseline. Second, an EC2 VM-only deployment in which all components are still deployed as VMs in EC2 but the components of both the front-end and logic layers use \Boxer. This deployment, \textit{EC2-only (w/ \Boxer)}, is used to measure the performance overhead of using \Boxer. Third, a mixed deployment where the front-end, and caching and storage layers are deployed as VMs, and the logic layer is deployed using Lambdas, \textit{EC2 + Lambda (w/ \Boxer)}. 

To measure the throughput and latency of the end-to-end system we use two workloads included in the DeathStarBench suite. A read workload that issues requests to read a user timeline in the social network, and a write workload that creates follow relationships between users. Both workloads are generated using the \texttt{wrk} \cite{wrk} tool which builds and issues requests to the front-end layer. The performance of both workloads (read and write) is reported separately as each workload stresses the network and \Boxer in a different way. The read workload mostly transfers data from the caching and storage layer (VMs), to the logic layer (VMs or Lambdas), and then to the front-end layer (VMs). The write workload operates in the opposite direction.

All experiments in this section were conducted in AWS Ohio (us-east-2) region. All VMs use a base Amazon Linux 2 \cite{amazon-linux}. For front-end, and caching and storage layers, we use t3a.micro instances due to the memory requirements of the services included in these layers. For logic layer, when deployed in VMs, we use t3a.nano instances. Lambdas are deployed using the Python3.7 runtime (which is used to launch \Boxer and the service binaries). Each Lambda is given 2048 MB of memory size (we experimentally determined that in us-east-2, the performance of a 2048 MB Lambda is similar to a t3a.nano VM instance).  

\subsection{Stateless Services in Lambdas}
We analyze the throughput and latency impact of using \Boxer to deploy the DeathStarBench social Network application and move the application logic layer to AWS Lambda. Figure \ref{fig:eval_deathstar} shows the results for both read and write workloads across the three different types of deployments. For each workload, we collect the average throughput and 90th percentile latency with an increasing load in the system.

Results show that \Boxer introduces only a low overhead. For the read workload, the EC2-only deployment becomes saturated at 3270 ops/s while the EC2-only (w/ \Boxer) becomes saturated at 3070 ops/s. For the same data points, the 90p latency of a single request for the EC2-only and EC2-only (w/ \Boxer) deployments are 3.18 ms and 5.07 ms, respectively. Note that these latencies are measured end-to-end and therefore include multiple internal microservice to microservice requests. The write workload demonstrates similar results. The EC2-only and EC2-only (w/ \Boxer) deployments become saturated at 1411 ops/s and 1294 ops/s, with latencies of 7.07 ms and 7.56 ms, respectively.

We use a similar analysis to measure the overhead of launching the logic layer services in AWS Lambda by comparing the EC2-only (w/ \Boxer) and EC2 + Lambda (w/ \Boxer) deployments.  Figure~\ref{fig:eval_deathstar} shows that for the read workload, the EC2 + Lambda (w/ \Boxer) deployment saturates at a 3556 ops/s with a 90p latency of 7 ms. For the write workload, the same deployment saturates at 1189 ops/s and with a 90p latency of 4.55 ms.

We conclude that using \Boxer incurs a small performance overhead arising from the interception and management of connections between the multiple services. Moving services to Lambda also incurs a small overhead due to the  different way CPU and Network are allocated to VMs and lambdas. One could further increase the memory budget given to lambdas to also increase their computational power and thus close the performance gap between EC2-only (w/ \Boxer) and EC2 + Lambda (w/ \Boxer).

\section{Dynamic Load Adaptation}\label{sec:dynamic}

We now show that \Boxer enables us to increase the scaling factor of stateless microservices running on VMs by leveraging the elasticity of serverless platforms to adapt to bursty loads. While in the previous experiments we deployed the logic layer of the DeathStarBench socialNetwork application entirely on Lambda or EC2, in this section we start by deploying all logic layer services on VMs. When the load increases, additional logic layer services are allocated to handle the increased load either on VMs or Lambdas. We compare two deployments: i) VM-based deployment that allocates extra VMs to handle increased load; ii) VM-based deployment that allocates lambdas to accommodate increased load. Our goal is to study the ability to adapt to a load burst.

Figure \ref{fig:eval_deathstar} (right-hand side) presents a throughput trace of both deployments. Throughput is measured using wrk \cite{wrk} by looking at how many requests the front-end layer can handle per second. After 30 seconds (dashed vertical line), a new set of VMs or lambdas are deployed to handle the increased load. After requesting a new set of VMs to join the logic layer (\textit{t=30s}), the system throughput starts increasing after 40 seconds (\textit{t=70s}) and only stabilizes (completely absorbs the burst) after 75 seconds (\textit{t=105s}). There are several reasons to this. First, VM launch time is approximately 30 seconds (i.e., until the VM is connected to the network). After the VM launches, we also register its IP in our nameserver (we use DNS round-robin to distribute the load). At this point, when NGINX (front-end layer) needs to open a new connection to serve a new client, it will try to resolve the logic layer component name through the nameserver. Because load distribution is performed per-connection and connections are kept open to serve multiple requests, adapting to a burst request takes dozens of seconds. The time interval between nodes becoming available and being fully utilized by NGINX could be reduced by forcing NGINX to recycle connections more frequently. This would, however, add overhead for creating more connections.

Using Lambdas to absorb load bursts can significantly reduce the time to add new logic layer services, and therefore, reduce the time to reach a steady throughput using all new nodes. Note that, although lambdas take less than one second to launch (compared to approximately 30 seconds for VMs), throughput starts increasing 8 seconds after the new lambdas are launched (\textit{t=38s}), and only stabilizes after 39 seconds (\textit{t=69s}). In summary, using Lambda to accommodate bursts reduces the time for new components start receiving requests by 5$\times$, and reduces the time to fully accommodate a burst by 1.9$\times$.

\section{Discussion}

We have shown that supporting direct networking among functions enables running off-the-shelf distributed applications, which were designed for traditional Linux execution environments, on serverless platforms. 
By decoupling the serverless resource management model from its event-driven programming model, \Boxer enables cloud users to benefit from the resource elasticity of serverless platforms without having to change the programming model of their applications.
The ability to run applications seamlessly between VM and serverless platforms allows users to leverage serverless platforms during bursty load periods and benefit from the lower cost per unit time of VM resources during steady load periods. 


\subsection{Opportunities}\label{discuss:opportunities}

Enabling direct networking among functions opens up many opportunities to address other key limitations of serverless platforms. For example, to overcome limitations on function execution time, a function that is about to reach the maximum supported execution time on the platform could spawn a follower function and send its state to the follower directly to transparently continue execution. 
For instance, Infinicache~\cite{InfiniCache} refreshes functions that cache data by having a function call itself and send the cached data to its newly spawned clone. With \Boxer, state can be sent through a direct communication channel, without relying on a proxy. 
Workflows can be programmed into functions so that a sequence of steps can execute automatically with functions spawning the next step and passing the necessary data without having to rely on an external orchestrator \cite{TxnOnServerless21,WorkflowOnServerless20}. In addition, scatter-gather interactions become possible and more efficient. This would considerably simplify running applications such as those considered in ExCamera~\cite{ExCamera} and gg~\cite{gg}, or removing a variety of work-arounds built to bypass the lack of function-to-function communication to implement stateful functions \cite{Cloudburst20,StatefulFunctions19}. It also enables implementing truly distributed data processing operators such as joins instead of using today's contrived solutions which need to communicate through storage~\cite{starling,Lambada}. Similar ideas apply to ML over serverless, which today is expensive due to the lack of communication \cite{ServerlessML21}. \Boxer can also be used to implement a form of work stealing among functions of a serverless applications, since functions could communicate directly with one another to request additional work if they are idle. 
This mechanism adds another dimension of elasticity to serverless computing, which is particular useful when the amount of work to be done is not easily determined upfront or may exhibit skew. 

Another key limitation of serverless platforms is that they do not allow users to optimize the placement of functions to improve performance~\cite{Berkeley-CACM}. By supporting direct networking between functions, cloud providers can get a complete picture of the communication patterns of an application and use this information to optimize function placement~\cite{sonic}. Understanding the communication patterns of serverless applications is difficult for providers to do today as functions are forced to communicate through remote storage systems or proxies. While we have shown how users can circumvent the networking limitations in today's serverless platforms with \Boxer, we hope that cloud providers will natively provide networking abstractions on serverless platforms. For example, providers could extend the serverless programming model to include communication collectives, which would allow them to optimize function placement policies for different communication patterns.

\subsection{Current limitations}\label{discuss:limitations}
\Boxer raises a number of issues that we intend to address in the future: 

\textbf{\Boxer control network topology:} Currently \Boxer establishes a fully connected TCP control topology between all participating \Boxer nodes. As the scale of deployment increase, this will likely be one of the bottlenecks.
We are investigating solutions such as indirection through other \Boxer nodes to route control messages, or using a datagram protocol to conserve the file descriptors.


\textbf{Non-blocking I/O:} \Boxer's transparent interposition layer does not support all non-blocking socket operations. Currently \Boxer will block a calling thread for the duration of connect call, which can degrade performance of applications and can break the semantics of some.


\textbf{Lambda execution environment:} Applications running in Lambda do not have access to the same exact environment that would be available in an EC2 VM. For example, the local file system might be read-only, there might be only one network device with a local IP, etc. \Boxer can circumvent these limitation and emulate an environment similar to the one available in VMs by further interception of the libc functions used to query local interfaces, local files, etc.

\textbf{Multiple lambdas behind a single NAT address:} \Boxer~currently does not support multiple lambdas deployed behind the same NAT address. When this happens, a single external IP address is visible and \Boxer~cannot distinguish between the two nodes. To support this scenario transparently to the user application, \Boxer~needs to virtualize the network address space.

\subsection{Truly general purpose short-lived datacenters}\label{discuss:general-purpose}

Besides networking, other limitations that serverless platforms impose today include limited resources per function invocation (e.g., up to 10 GB of memory and 6 vCPUs in AWS Lambda), limited execution time for each function (e.g., 15 minutes in AWS Lambda), and the lack of support for heterogeneous hardware such as GPUs. Cloud providers have been steadily increasing resource limits and function execution time limits. We expect this trend to continue as serverless computing becomes increasingly popular for a broad range of applications. In particular, supporting higher resource limits per function gives developers the option to scale \textit{up} in addition to scale \textit{out} their application to achieve higher performance. Both of these scaling dimensions are important for general datacenter computing. Adding support for running serverless functions on heterogeneous hardware resources, such as accelerators, will also become increasingly important as applications such as machine learning jobs, which rely on GPUs and ASICs, can also benefit from the elasticity, fine-grain billing, and higher level of abstraction to the cloud that serverless computing offers.

\section{Conclusion}

We presented \Boxer, a system that transparently provides direct communication between serverless functions. \Boxer uses a NAT traversal mechanism to enable serverless functions to accept incoming traffic from external sources. By leveraging dynamic linking to intercept function calls, \Boxer provides direct networking functionality for serverless functions through the standard socket networking API and requires no modifications to Linux TCP-based datacenter applications. With \Boxer, latency sensitive applications such as microservices can be run unmodified on AWS Lambda. Our system presents a major step towards treating serverless platforms as short-lived, instant, general purpose datacenters and allows a broader range of applications to benefit from the elasticity, fine-grain billing, and high level of abstraction to the cloud that serverless platforms offer.

\printbibliography

@inproceedings{Boxer-CIDR21,
author={Michal Wawrzoniak and Ingo Müller and Rodrigo Bruno and Gustavo Alonso},
title={Boxer: Data Analytics on Network-enabled Serverless Platforms},
Booktitle={{CIDR}},
year={2021},
}

@article{Berkeley-CACM,
author = {Schleier-Smith, Johann and Sreekanti, Vikram and Khandelwal, Anurag and Carreira, Joao and Yadwadkar, Neeraja J. and Popa, Raluca Ada and Gonzalez, Joseph E. and Stoica, Ion and Patterson, David A.},
title = {What Serverless Computing is and Should Become: The next Phase of Cloud Computing},
year = {2021},
volume = {64},
number = {5},
journal = {Commun. ACM},
month = apr,
pages = {76–84},
}

@inproceedings{Particle20,
author = {Thomas, Shelby and Ao, Lixiang and Voelker, Geoffrey M. and Porter, George},
title = {Particle: Ephemeral Endpoints for Serverless Networking},
year = {2020},
booktitle = {Proceedings of the 11th ACM Symposium on Cloud Computing},
pages = {16–29}
}

@inproceedings{Nightcore21,
author = {Jia, Zhipeng and Witchel, Emmett},
title = {Nightcore: Efficient and Scalable Serverless Computing for Latency-Sensitive, Interactive Microservices},
year = {2021},
booktitle = {Proceedings of the 26th ACM International Conference on Architectural Support for Programming Languages and Operating Systems},
pages = {152–166}
}

@inproceedings{WorkflowOnServerless20,
  author    = {Haoran Zhang and
               Adney Cardoza and
               Peter Baile Chen and
               Sebastian Angel and
               Vincent Liu},
  title     = {Fault-tolerant and transactional stateful serverless workflows},
  booktitle = {14th {USENIX} Symposium on Operating Systems Design and Implementation,
               {OSDI} 2020, Virtual Event, November 4-6, 2020},
  pages     = {1187--1204},
  year      = {2020}
}

@article{Cloudburst20,
author = {Sreekanti, Vikram and Wu, Chenggang and Lin, Xiayue Charles and Schleier-Smith, Johann and Gonzalez, Joseph E. and Hellerstein, Joseph M. and Tumanov, Alexey},
title = {Cloudburst: Stateful Functions-as-a-Service},
year = {2020},
volume = {13},
number = {12},
journal = {Proc. VLDB Endow.},
pages = {2438–2452},
}

@inproceedings{TxnOnServerless21,
author = {de Heus, Martijn and Psarakis, Kyriakos and Fragkoulis, Marios and Katsifodimos, Asterios},
title = {Distributed Transactions on Serverless Stateful Functions},
year = {2021},
booktitle = {Proceedings of the 15th ACM International Conference on Distributed and Event-Based Systems},
pages = {31–42},
}

@article{StatefulFunctions19,
author = {Akhter, Adil and Fragkoulis, Marios and Katsifodimos, Asterios},
title = {Stateful Functions as a Service in Action},
year = {2019},
volume = {12},
number = {12},
journal = {Proc. VLDB Endow.},
pages = {1890–1893}
}

@inproceedings{ServerlessML21,
author = {Jiang, Jiawei and Gan, Shaoduo and Liu, Yue and Wang, Fanlin and Alonso, Gustavo and Klimovic, Ana and Singla, Ankit and Wu, Wentao and Zhang, Ce},
title = {Towards Demystifying Serverless Machine Learning Training},
year = {2021},
booktitle = {Proceedings of the 2021 International Conference on Management of Data},
pages = {857–871}
}

@article{RiseofServerless19,
author = {Castro, Paul and Ishakian, Vatche and Muthusamy, Vinod and Slominski, Aleksander},
title = {The Rise of Serverless Computing},
year = {2019},
volume = {62},
number = {12},
journal = {Commun. ACM},
pages = {44–54}
}

@inproceedings{InfiniCache,
  author    = {Ao Wang and
               Jingyuan Zhang and
               Xiaolong Ma and
               Ali Anwar and
               Lukas Rupprecht and
               Dimitrios Skourtis and
               Vasily Tarasov and
               Feng Yan and
               Yue Cheng},
  title     = {InfiniCache: Exploiting Ephemeral Serverless Functions to Build a
               Cost-Effective Memory Cache},
  booktitle = {{USENIX} {FAST}},
  year      = {2020},
}

@inproceedings{HellersteinCIDR19,
  author    = {Joseph M. Hellerstein and
               Jose M. Faleiro and
               Joseph Gonzalez and
               Johann Schleier{-}Smith and
               Vikram Sreekanti and
               Alexey Tumanov and
               Chenggang Wu},
  title     = {Serverless Computing: One Step Forward, Two Steps Back},
  booktitle = {{CIDR}},
  year      = {2019},
  }

@inproceedings{Anna18,
  author    = {Chenggang Wu and
               Jose M. Faleiro and
               Yihan Lin and
               Joseph M. Hellerstein},
  title     = {{Anna: {A} {KVS} for Any Scale}},
  booktitle = {ICDE},
  year      = {2018},
  }

@inproceedings{Ford,
  author    = {Bryan Ford and
               Pyda Srisuresh and
               Dan Kegel},
  title     = {Peer-to-Peer Communication Across Network Address Translators},
  booktitle = {{USENIX} {ATC}},
  year      = {2005},
}

@inproceedings{Eppinger,
  author    = {Eppinger J.L.},
  title     = {TCP Connections for P2P Apps: A Software Approach
to Solving the NAT Problem},
  booktitle = {Carnegie Mellon University, Tech. Rep, ISRI-05-104},
  year      = {2005},
}

@inproceedings{DeathStarBench,
author = {Gan, Yu and Zhang, Yanqi and Cheng, Dailun and Shetty, Ankitha and Rathi, Priyal and Katarki, Nayan and Bruno, Ariana and Hu, Justin and Ritchken, Brian and Jackson, Brendon and Hu, Kelvin and Pancholi, Meghna and He, Yuan and Clancy, Brett and Colen, Chris and Wen, Fukang and Leung, Catherine and Wang, Siyuan and Zaruvinsky, Leon and Espinosa, Mateo and Lin, Rick and Liu, Zhongling and Padilla, Jake and Delimitrou, Christina},
title = {An Open-Source Benchmark Suite for Microservices and Their Hardware-Software Implications for Cloud \& Edge Systems},
year = {2019},
booktitle = {{ASPLOS}},
}

@INPROCEEDINGS{UDT,
  author={ {Yunhong Gu} and  {Xinwei Hong} and R. L. {Grossman}},
  booktitle={SC '04}, 
  title={Experiences in Design and Implementation of a High Performance Transport Protocol}, 
  year={2004},
  }

@inproceedings {Locus,
author = {Qifan Pu and Shivaram Venkataraman and Ion Stoica},
title = {Shuffling, Fast and Slow: Scalable Analytics on Serverless Infrastructure},
booktitle = {{NSDI} 19},
year = {2019},
}

@inproceedings {ExCamera,
author = {Sadjad Fouladi and Riad S. Wahby and Brennan Shacklett and Karthikeyan Vasuki Balasubramaniam and William Zeng and Rahul Bhalerao and Anirudh Sivaraman and George Porter and Keith Winstein},
title = {Encoding, Fast and Slow: Low-Latency Video Processing Using Thousands of Tiny Threads},
booktitle = {{NSDI}},
year = {2017},
}

@inproceedings {gg,
author = {Sadjad Fouladi and Francisco Romero and Dan Iter and Qian Li and Shuvo Chatterjee and Christos Kozyrakis and Matei Zaharia and Keith Winstein},
title = {From Laptop to Lambda: Outsourcing Everyday Jobs to Thousands of Transient Functional Containers},
booktitle = {{USENIX} {ATC}},
year = {2019},
}

@inproceedings{Lambada,
author = {Müller, Ingo and Marroquín, Renato and Alonso, Gustavo},
title = {Lambada: Interactive Data Analytics on Cold Data Using Serverless Cloud Infrastructure},
year = {2020},
booktitle = {SIGMOD},
}

@misc{iperf3,
title = {{iperf3}},
url = {https://software.es.net/iperf/},
urldate ={2020-12-10}
}

@misc{wrk,
title = {{wrk - a HTTP benchmarking tool}},
url = {https://github.com/wg/wrk},
urldate ={2021-10-09}
}

@misc{awslambda,
title = {{AWS Lambda}},
url = {https://aws.amazon.com/lambda},
urldate = {2020-08-17}
}

@misc{Amazon-Lambda-reservation-system,
title = {{Managing AWS Lambda Function Concurrency}},
url = {https://aws.amazon.com/blogs/compute/managing-aws-lambda-function-concurrency/},
urldate = {2020-08-28}
}

@misc{S3,
title = {{Amazon S3}},
url = {https://aws.amazon.com/s3},
urldate = {2020-08-17}
}

@misc{azurefunctions,
title = {{Microsoft Azure Functions}},
url = {https://azure.microsoft.com/en-us/services/functions},
urldate = {2020-08-17}
}

@misc{googlefunctions,
title = {{Google Cloud Functions}},
url = {https://cloud.google.com/functions},
urldate = {2020-12-1}
}

@misc{aws-1ms-billing,
title = {{AWS Lambda changes duration billing granularity from 100ms down to 1ms}},
url = {https://aws.amazon.com/about-aws/whats-new/2020/12/aws-lambda-changes-duration-billing-granularity-from-100ms-to-1ms/},
urldate = {2020-12-1}
}

@misc{aws-lambda-containers,
title = {{AWS Lambda now supports container images as a packaging format}},
url = {https://aws.amazon.com/about-aws/whats-new/2020/12/aws-lambda-now-supports-container-images-as-a-packaging-format/},
urldate = {2020-12-1}
}

@misc{aws-FAQs,
title = {{AWS Lambda FAQs}},
url = {https://aws.amazon.com/lambda/faqs/},
urldate = {2020-12-1}
}

@misc{amazon-linux,
title = {{Amazon Linux 2}},
url = {https://aws.amazon.com/amazon-linux-2/},
urldate = {2020-12-10}
}

@misc{serverlessnet,
title = {{Serverless Networking SDK}},
url = {http://networkingclients.serverlesstech.net/},
urldate = {2020-08-17}
}

@article{anna,
author = {Wu, Chenggang and Sreekanti, Vikram and Hellerstein, Joseph M.},
title = {Autoscaling Tiered Cloud Storage in Anna},
year = {2019},
journal = {{PVLDB}},
}

@inproceedings{pocket,
author = {Klimovic, Ana and Wang, Yawen and Stuedi, Patrick and Trivedi, Animesh and Pfefferle, Jonas and Kozyrakis, Christos},
title = {Pocket: Elastic Ephemeral Storage for Serverless Analytics},
year = {2018},
booktitle = {{OSDI}},
pages = {427–444},
numpages = {18},
}

@inproceedings{starling,
author = {Perron, Matthew and Castro Fernandez, Raul and DeWitt, David and Madden, Samuel},
title = {Starling: A Scalable Query Engine on Cloud Functions},
year = {2020},
booktitle = {SIGMOD},
}

@inproceedings {sonic,
author = {Ashraf Mahgoub and Karthick Shankar and Subrata Mitra and Ana Klimovic and Somali Chaterji and Saurabh Bagchi},
title = {{SONIC}: Application-aware Data Passing for Chained Serverless Applications},
booktitle = {{USENIX} Annual Technical Conference ({USENIX} {ATC} 21)},
year = {2021},
}

\end{document}